\documentclass[ALICE,manyauthors]{cernphprep}
\usepackage[%
            final,
            color,
            eps,
            ]{style}

\begin{document}
\begin{titlepage}
  \PHyear{2015}
  \PHnumber{257}
  \PHdate{16 September} 
  \title{Centrality evolution of the charged--particle pseudorapidity density
    over a broad pseudorapidity range in Pb--Pb collisions at $\sNN*$}
  \ShortTitle{Centrality evolution of the charged-particle
    pseudorapidity density in Pb-Pb}%
  \Collaboration{ALICE Collaboration\thanks{See
      Appendix~\ref{app:collab} for the list of collaboration
      members}}%
  \ShortAuthor{ALICE Collaboration}%
  \abstract{ \noindent 
The centrality dependence of the charged--particle pseudorapidity density
measured with ALICE in Pb--Pb collisions at $\sNN$ over a broad
pseudorapidity range is presented.  
This Letter extends the previous results reported by ALICE to more
peripheral collisions. 
No strong change of the overall shape of charged--particle pseudorapidity
density distributions with centrality is observed, 
and when normalised to the number of participating nucleons in the
collisions, the evolution over pseudorapidity with centrality is likewise
small.  
The broad pseudorapidity range ($-3.5<\eta<5$) allows precise estimates of the
total number of produced charged particles which we find to range from
$162\pm22\text{(syst.)}$ to $17170\pm770\text{(syst.)}$ in
\centRange{80}{90} and \centRange{0}{5} central collisions,
respectively. 
The total charged--particle multiplicity is seen to approximately
scale with the number of participating nucleons in the collision.
This suggests that hard contributions to the charged--particle
multiplicity are limited.  
The results are compared to models which describe $\ndndeta$ at
mid--rapidity in the most central Pb--Pb collisions and it is found that
these models do not capture all features of the distributions. 


} \MarkSvnVersion{}
\end{titlepage}
\addtocounter{page}{-1}

\section{Introduction}

The measurement of the charged--particle pseudorapidity ($\eta$) density
distribution in heavy--ion collisions provides insight into the
dominant particle production mechanisms, such as parton
fragmentation~\cite{Gribov:1984tu} and the observed phenomenon of
limiting fragmentation~\cite{PhysRevC.83.024913}. The unique
capability of ALICE to perform such measurements from large to small
overlaps of the colliding nuclei over a broad pseudorapidity range allows
for significant additional information to be extracted e.g., the total
number of charged particles and the evolution of the distributions
with centrality.

The charged--particle pseudorapidity density ($\ndndeta$) \textit{per se} does
not provide immediate understanding of the particle production
mechanism, but as a benchmark tool for comparing models it is
indispensable.  Various models
\cite{Wang:1991hta,Lin:2004en,Pierog:2013ria} make different
assumptions on how particles are produced in heavy--ion collisions
resulting in very different charged--particle pseudorapidity density
distributions --- both in terms of scale and shape.  Models may, for
example, incorporate different schemes for the hadronisation of the
produced quarks and gluons which leads to very different pseudorapidity
distributions of the charged particles.

The ALICE collaboration has previously reported results on the
charged--particle pseudorapidity density in the \centRange{0}{30} most central
Pb--Pb collisions at $\sNN{}$ over a wide pseudorapidity
range~\cite{Abbas:2013bpa}, and in the 80\% most central collisions at
mid--rapidity ($\eta\approx0$) only~\cite{Aamodt:2010cz}. The ATLAS
collaboration has reported on the charged--particle pseudorapidity density in
the 80\% most central events in a limited pseudorapidity range of
$|\eta|<2$~\cite{ATLAS:2011ag}.  Similarly, the CMS collaboration has
reported on the same measurements in the 90\% most central events at
$\eta\approx0$, and for selected centralities up to
$|\eta|<2$~\cite{Chatrchyan:2011pb}.

In this Letter we present the primary charged--particle pseudorapidity density
dependence on the event centrality from mid--central
(\centRange{30}{40}) to peripheral (\centRange{80}{90}) collisions
over a broad pseudorapidity range to complement results previously reported by
ALICE in the \centRange{0}{30} centrality range.  Unlike previous
\cite{Abbas:2013bpa}, in the forward regions where the signal is
dominated by secondary particles produced in the surrounding material,
we use a data-–driven correction to extract the primary
charged--particle density. 

Primary charged particles are defined as prompt charged particles
produced in the collision, including their decay products, but
excluding products of weak decays of muons and light flavour
hadrons. 
Secondary charged particles are all other particles observed in the
experiment e.g., particles produced through interactions with material
and products of weak decays.

In the following section, the experimental set--up will be briefly
described. \secref*{sec:data} outlines analysis procedures and
describes a data--driven method to isolate the number of primary
charged particles from the secondary particle background at large
pseudorapidity.  Systematic uncertainties are discussed in
\secref{sec:sysuncer}. %
In \secref{sec:results}, the resultant charged--particle
pseudorapidity density distributions are presented along with their
evolution with centrality. Furthermore we extract from the measured
$\ndndeta$ distributions the total number of charged particles as a
function of the number of participating nucleons. %
We finally compare the measured charged--particle pseudorapidity
density to a number of model predictions before concluding in
\secref{sec:concl}.



\section{Experimental setup}
\label{sec:exper}

A detailed description of ALICE can be found
elsewhere~\cite{Aamodt:2008zz,Abelev:2014ffa}. In the following we
briefly describe the detectors relevant to this analysis.

The Silicon Pixel Detector (SPD) is the inner--most detector of ALICE.
The SPD consists of two cylindrical layers of $9.8\times10^6$
silicon--pixels possessing binary read--out. It provides a measurement
of charged particles over $|\eta|<2$ using so-called \emph{tracklets}
--- a combination of hits on each of the two layers (1 and 2)
consistent with a track originating from the interaction point.
Possible combinations of hits not consistent with primary particles
can be removed from the analysis, with only a small (a few $\%$)
residual correction for secondary particles derived from simulations.
The SPD also provides a measurement, by combining hits on its two
layers, of the offset with respect to the interaction point, where the
collisions occurred.
$\text{\textbf{IP}}=(0,0,0)$ is at the centre of the ALICE coordinate
system, and $\IPz$ is the offset along the beam axis. Finally, a
hardware logical \emph{or} of hits in each of the two layers provides
a trigger for ALICE.

The Forward Multiplicity Detector (FMD) is a silicon strip detector
with $51\,200$ individual read--out channels recording the energy
deposited by particles traversing the detector.  It consists of three
sub--detectors FMD1, 2, and 3, placed approximately $\unit[320]{cm}$,
$\unit[79]{cm}$ and $\unit[-69]{cm}$ along the beam line,
respectively.  FMD1 consists of one inner type ring (1i), while both
FMD2 and 3 consist of inner (2i,3i) and outer type rings (2o,3o).  The
rings have almost full coverage in azimuth ($\varphi$), and high
granularity in the radial ($\eta$) direction (see
\tabref{tab:detectors}).

The V0 is the most forward of the three detectors used in this
analysis.  It consists of two sub-detectors: \mbox{V0-A} and
\mbox{V0-C} placed at approximately $\unit[333]{cm}$ and
$\unit[-90]{cm}$ along the beam line, respectively.  Each of the
sub--detectors are made up of scintillator tiles with a high timing
resolution.  While the V0 provides pulse--height measurements, the
energy--loss resolution is not fine enough to do an independent
charged particle measurement.  In previous measurements, using
so--called satellite--main collisions (see \secref{sec:data}), one
could match the V0 amplitude to the SPD measurements to obtain a
relative measurement of the number of charged particles.  However, for
collisions at $|\IPz|<\unit[15]{cm}$ no such matching is possible, and
the V0 is therefore not used to provide a measurement of the number of
charged particles in this analysis.  The detector is used, in an
inclusive logical \emph{or} with the SPD, for triggering ALICE and to
provide a measure of the event centrality~\cite{Aamodt:2010cz}.

Details on the coverage, resolution, and segmentation of the three
used detectors are given in \tabref{tab:detectors}.

\def\rangeTo{to}
\begin{table}[htbp]
  \centering
  \small
  \begin{tabular}[T]{%
    r@{}lccc}
    \hline
    \multicolumn{2}{c}{\textbf{Detector}}
    & ${\delta r\varphi}$ 
    & ${\delta z}$ 
    & ${\eta}\ \text{range}$\\
    \hline 
    SPD 
    & 1
    & $\unit[12]{\upmu{}m}$
    & $\unit[100]{\upmu{}m}$ 
    & $-2.0$ \rangeTo{} $\phantom{-}2.0$ \\
    & 2
    & $\unit[12]{\upmu{}m}$
    & $\unit[100]{\upmu{}m}$ 
    & $-1.4$ \rangeTo{} $\phantom{-}1.4$ \\
    \hline 
    \multicolumn{2}{c}{\textbf{Detector}}
    & ${\Delta \varphi}$
    & ${\Delta r}$
    & ${\eta}\ \text{range}$\\
    \hline 
    FMD 
    & 1i 
    & $\unit[18]{{}^\circ}$ 
    & $\unit[254]{\upmu{}m}$ 
    & $\phantom{-}3.7$ \rangeTo{} $\phantom{-}5.0$\\
    & 2i 
    &  $\unit[18]{{}^\circ}$ 
    & $\unit[254]{\upmu{}m}$ 
    & $\phantom{-}2.3$ \rangeTo{} $\phantom{-}3.7$\\
    & 2o 
    & $\unit[\phantom{1}9]{{}^\circ}$ 
    & $\unit[508]{\upmu{}m}$ 
    & $\phantom{-}1.7$ \rangeTo{} $\phantom{-}2.3$ \\
    & 3o 
    & $\unit[\phantom{1}9]{{}^\circ}$ 
    & $\unit[508]{\upmu{}m}$ 
    & $-2.3$ \rangeTo{} $-1.7$\\
    & 3i 
    &  $\unit[18]{{}^\circ}$ 
    & $\unit[254]{\upmu{}m}$ 
    & $ -3.4$ \rangeTo{} $-2.0$ \\
    \hline 
    V0 
    & -A 
    & $\unit[45]{{}^\circ}$ 
    & $34$ \rangeTo{} $\unit[186]{mm}$ 
    & $\phantom{-}2.8$ \rangeTo{} $\phantom{-}5.1$ \\
    & -C 
    & $\unit[45]{{}^\circ}$ 
    & $26$ \rangeTo{} $\unit[127]{mm}$ 
    & $-3.7$ \rangeTo{} $-1.7$ \\
    \hline 
  \end{tabular}
  \caption{Overview of the resolution ($\delta$),
    segmentation ($\Delta$), and coverage of the detectors used in the
    analysis. The `A' side corresponds to $z>0$, while the `C' side
    corresponds to $z<0$.  The $\eta$ range is specified for
    collisions with $\IPz=0$.} 
  \label{tab:detectors}
\end{table}

\section{Data sample and analysis method}
\label{sec:data}

The results presented in this paper are based on Pb--Pb collision
data at $\sNN$ taken by ALICE in 2010.  About $100\,000$ events with a
minimum bias trigger requirement \cite{Aamodt:2010cz} were analysed in
the centrality range from $0\%$ to $90\%$.  The data was collected
over roughly 30 minutes where the experimental conditions did not
change.

The standard ALICE event selection~\cite{Aamodt:2010pb} and centrality
estimator based on the V0--amplitude are used in this
analysis~\cite{Abelev:2013qoq}.  We include here the
\centRange{80}{90} centrality class which was not present in the
previous results~\cite{Aamodt:2010cz}.  As discussed
elsewhere~\cite{Abelev:2013qoq}, the \centRange{90}{100} centrality
class has substantial contributions from QED processes and is therefore not
included in this Letter.

Results in the mid--rapidity region ($|\eta|<2$) are obtained from a
tracklet analysis using the two layers of the SPD as mentioned in
\secref{sec:exper}.  The analysis method and data used are identical
to what has previously been
presented~\cite{Abbas:2013bpa,Aamodt:2010cz}.

The measurements in the forward region ($|\eta|>2$) are provided by the
FMD.  The FMD records the full energy deposition of charged particles
that impinge on the detector.  Since all charged particles that hit
the FMD are boosted in the laboratory frame, the detection efficiency
is close to 100\% for all momenta.  As reported
earlier~\cite{Abbas:2013bpa}, the main challenge in measuring the
number of charged primary particles in this region, is the large
background of secondary particles produced in the surrounding
material.  Due to the complexity and the limited knowledge of the
material distribution of support structures away from the central
barrel, it has not been possible to adequately describe (on the few
\%--level) the generation of secondary particles in the forward
directions within the precision of the current simulation of the
ALICE apparatus. 

A suitable means to extract the number of primary charged particles
was found by utilising collisions between so--called `satellite'
bunches and main bunches offset in intervals of $\unit[37.5]{cm}$
along the beam--line.  Satellite bunches are caused by the so--called
debunching effect \cite{Welsch:2012zzf}.  A small fraction of the beam
can be captured in unwanted RF buckets, due the way beams are injected
into the accelerator, and create these satellite bunches spaced by
$\unit[2.5]{ns}$.  Collisions between satellite and main bunches can
cause instabilities in the beam, and the LHC has taken steps to reduce
the number of these kinds of collisions.  ALICE has therefore not
recorded collisions between satellite and main bunches before or after
the Pb--Pb run of 2010.  In satellite--main collisions the background
of secondary particles was much smaller and much better understood
since significantly less detector material shadows the forward
detectors \cite{Abbas:2013bpa}.   

A study utilising these satellite--main collisions led to the
publication of the measurement of the charged--particle pseudorapidity
density in the 30\% most central events over
$|\eta|<5$~\cite{Abbas:2013bpa}.  The study was limited in centrality
reach by the need to use the Zero--Degree Calorimeter (ZDC) for the
centrality estimation for collisions between satellite and main
bunches.
The ZDC measures the energy of spectator (non--interacting) nucleons
with two components: one measures protons and the other measures
neutrons. The ZDC was located at about \unit[114]{m} from the
interaction point on either side of the
experiment~\cite{Aamodt:2008zz}, and was therefore ideally suited for
that study. The centrality determination capability of the ZDC is
however limited to the $30\%$ most central
collisions~\cite{Abelev:2013qoq}.

For centralities larger than $30\%$ the V0 amplitude is used as the
centrality estimator, which is available only for collisions at
$|\IPz|<\unit[15]{cm}$ --- the so--called nominal interaction point
corresponding to main bunches of one beam colliding with main bunches
of the other beam.

To extend the centrality reach of the $\ndndeta{}$ measurement, a
\emph{data--driven} correction for the number of secondaries impinging
on the FMD has been implemented.  For each centrality class $C$, we
form the ratio
\begin{equation}
  \label{eq:empcor}
  E_C(\eta) = \frac{\ndndeta[C,\text{inclusive,nominal}]}{%
    \ndndeta[C,\text{primary,satellite}]}\quad.
\end{equation}
That is, the ratio of the \emph{measured} \emph{inclusive}
charged--particle density from main--main collisions
($|\IPz|<\unit[10]{cm}$) provided by the FMD to the \emph{primary}
charged--particle density from satellite--main
collisions~\cite{Abbas:2013bpa}.  Here, `inclusive' denotes primary
\emph{and} secondary charged particles i.e., no correction was applied
to account for secondary particles impinging on the FMD.

Note, that the correction is formed bin--by--bin in pseudorapidity, so that
the pseudorapidity is the same for both the numerator and denominator.
However, the numerator and denominator differ in the offset along the
beam line of origin of the measured particles: For the numerator the
origin lies within the nominal interaction region, while for the
denominator the origin was offset by multiples of \unit[37.5]{cm}.

This ratio is obtained separately for all previously published
centrality classes: \centRange{0}{5}, \centRange{5}{10},
\centRange{10}{20} and \centRange{20}{30}.  The variation of
$E_c$ for different centralities is small ($<1\%$, much smaller than
the precision of the measurements). The weighted average
\begin{equation}
  \label{eq:empcor2}
  E(\eta) = \frac{\sum_C \Delta C E_{C}(\eta)}{\sum_C \Delta C}\quad,
\end{equation}
is used as a global correction to obtain the primary charged--particle
pseudorapidity density
\begin{equation}
  \label{eq:dndeta}
  \ndndeta[X,\text{primary}] = 
  \frac{1}{E(\eta)}\ndndeta[X,\text{inclusive,nominal}]\quad,
\end{equation}
where $X$ stands for an event selection e.g., a centrality range.

The simulation--based correction $S(\eta)$ for secondary particles to
the charged--particle pseudorapidity density in the forward directions is
given by
\begin{equation}
  \label{eq:secmap}
S(\eta) =
\frac{N_{\text{inclusive,FMD}}(\eta)}{N_{\text{primary,generated}}(\eta)}\quad,
\end{equation}
where $N_{\text{inclusive,FMD}}$ is the number of primary \emph{and}
secondary particles impinging on the FMD --- as given by the track
propagation of the simulation, and $N_{\text{primary,generated}}$ is
the number of generated primary particles at a given
pseudorapidity. Complete detector--simulation studies show that three
effects can contribute to the generation of secondaries, and hence the
value of $S(\eta)$.  These three effects are: material in which
secondaries are produced, the transverse momentum ($\pT$) distribution
and particle composition of the generated particles, and lastly the
total number of produced particles.  Of these three the material is by
far the dominant effect, while the $\pT$ and particle composition only
effects $S(\eta)$ on the few percent level.  The total number of
generated particles has a negligible effect on $S(\eta)$.  That is,
the material surrounding the detectors amplifies the primary--particle
signal through particle production by a constant factor that first and
foremost depends on the amount of material itself, and only
secondarily on the $\pT$ and particle composition of the generated
primary particles.

To estimate how much $E_{\text{C}}(\eta)$ itself would have changed if
another system or centrality range was used to calculate the
correction, $S(\eta)$ is analysed from simulations with various
collision systems and energies.  We find that, even for large
variations in particle composition and $\pT$ distributions, $S(\eta)$
only varies by up to $5\%$.  Re--weighting the particle composition
and $\pT$ distributions from the various systems to match produces
consistent values of $S(\eta)$ ensuring that the $5\%$ variations
found were only due to particle composition and $\pT$ distributions
differences.  This uncertainty is applied to $E(\eta)$ to account for
all reasonable variations of the particle composition and $\pT$
distributions, which cannot be measured in the forward regions of
ALICE.

\begin{figure}[htbp]
  \centering
  \figinput[\linewidth]{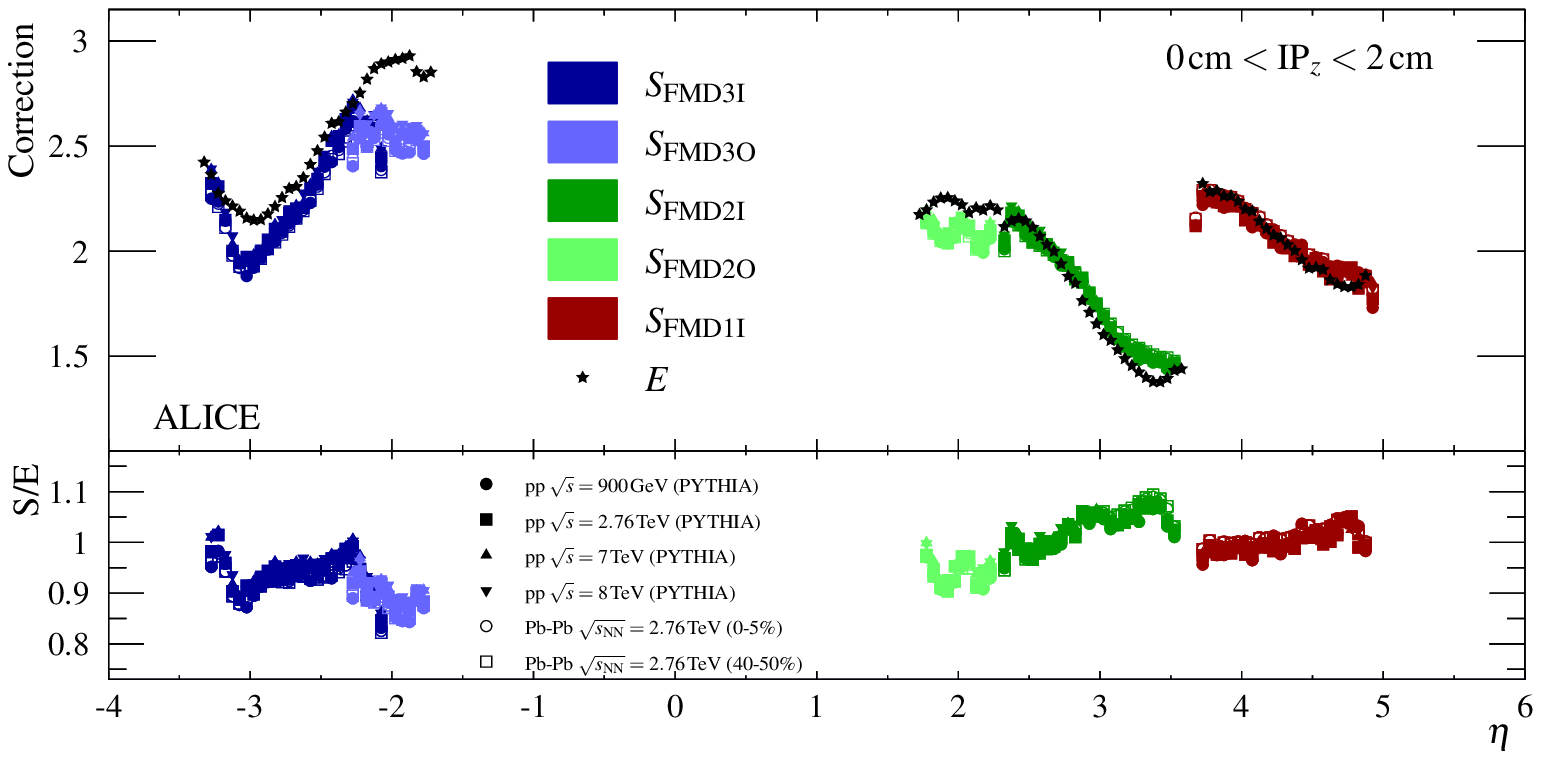}
  \input{figs/caps/SecmapEmp.cap}
  \label{fig:SecmapEmp}
\end{figure}

\figref*{fig:SecmapEmp} shows the comparison of the data driven
correction $E(\eta)$ to the simulation--based correction $S(\eta)$
from \textsc{Pythia}~\cite{Sjostrand:2006za} (pp) and a
parameterisation of the available ALICE
results~\cite{Abelev:2014pua,Abelev:2013vea} for Pb--Pb
collisions. The simulated collisions are for two distinct systems and span over almost an order of
magnitude in collision energy.  The
total number of produced particles in these simulations span five orders of
magnitude, and no dependence of $S(\eta)$ on
charged--particle multiplicity is observed.

By comparing $E(\eta)$ to $S(\eta)$ from simulations, one finds a good
correspondence between the two corrections \emph{except} in regions
where the material description in the simulations is known to be
inadequate. This, together with the fact that the numerator and
denominator of \eqref{eq:empcor} measure the same physical process,
but differ foremost in the material traversed by the primary
particles, and hence the number of secondary particles observed,
implies that the correction $E(\eta)$ is universal.  That is,
\eqref{eq:dndeta} is applicable for \emph{any} event selection $X$ in
any collision system or at any collision energy, where the produced
multiplicity, $\pT$ distributions, and particle composition is close
to the range of the simulated systems used to study $S(\eta)$.

Note, for the previously published results~\cite{Abbas:2013bpa}, which
used satellite--main collisions, the simulation--based approach for
correcting for secondary particles i.e., applying $S(\eta)$ directly,
was valid.  As mentioned above, in satellite--main collisions, the
particles that impinge on the FMD traverse far less and better
described material in the simulation of the ALICE apparatus.  The use
of a simulation--based correction for secondary particles was in that
analysis cross--checked by comparing to and combining with
measurements from the V0 and SPD~\cite{Abbas:2013bpa}. Despite
concerted efforts to improve the simulations by the Collaboration it
has not been possible to achieve the same accuracy in $S(\eta)$ for
main--main collisions.

Finally, the effect of variation of the location of the primary
interaction point on $E(\eta)$ was studied.  It was found, that the
effect is negligible, given that the distribution of $\IPz$ are
similar between the numerator of \eqref{eq:empcor} and right--hand
side of \eqref{eq:dndeta}, as was the case in this analysis. 

The method used in this analysis to extract the inclusive number of
charged particles from the FMD is the same as for previous published
results~\cite{Abbas:2013bpa}, \emph{except} that the data--driven
correction $E(\eta)$ --- rather than a simulation--based one $S(\eta)$
--- is used to correct for secondary particles.



\section{Systematic uncertainties} 
\label{sec:sysuncer}

\tabref*{tab:sysuncer} summarises the systematic uncertainties
of this analysis.  The common systematic uncertainty from the
centrality selection is correlated across $\eta$ and detailed
elsewhere~\cite{Abelev:2013qoq}.   
 
For the SPD measurements, the systematic uncertainties are the same as
for the previously published mid--rapidity result~\cite{Aamodt:2010cz},
except for a contribution from the correction due to the larger
acceptance used in this analysis.  This uncertainty stems from the
range of $\IPz$ used in the analysis (here $|\IPz|<\unit[15]{cm}$).
At larger absolute values of $\IPz$ the acceptance correction for the
SPD tracklets grows, and the uncertainty with it, being therefore
$\eta$--dependent and largest at $|\eta|\approx2$.

The various sources of systematic uncertainties for the FMD
measurements are detailed elsewhere~\cite{Abbas:2013bpa}, but will be
expanded upon in the following since some values have changed due to
better understanding of the detector response.

In the analysis, three $\eta$--dependent thresholds are used. The
values for these thresholds are obtained by fitting a convoluted
Landau--Gauss distribution~\cite{Hancock:1983ry} to the energy loss
spectrum measured by the FMD in a given $\eta$ range.  The
uncertainties associated with these thresholds are detailed below. 

A charged particle traversing the FMD can deposit energy in more than
one element i.e., strip, of the detector. Therefore it is necessary to
re--combine two signals to get the single charged--particle energy
loss in those cases.  This \systUncName{recombination} depends on a
lower threshold for accepting a signal, and an upper threshold to
consider a signal as isolated i.e., all energy is deposited in a
single strip.  The systematic uncertainties from the recombination of
signals are found by varying the lower and upper threshold values
within bounds of the energy loss fits and by simulation studies.

To calculate the inclusive number of charged particles, a statistical
approach is used~\cite{Abbas:2013bpa}.  The strips of the FMD are
divided into regions, and the number of empty strips is compared to
the total number of strips in a given region.  Strips with a signal
below a given \systUncName{threshold} are considered empty.  The
threshold was varied within bounds of the energy loss fits and
investigated in simulation studies to obtain the systematic
uncertainty.

The data--driven correction for \systUncName{secondary particles}
defined in \eqref{eq:empcor2} is derived from the previously published
results, and as such contains contributions from the systematic
uncertainties of those results~\cite{Abbas:2013bpa}.  Factoring out
common correlated uncertainties e.g., the contribution from the
centrality determination, we find a contribution of $4.7\%$ from the
previously published results.  By studying the variation of the
numerator of \eqref{eq:empcor} under different experimental conditions
e.g., different data--taking periods, and adding the variance in
quadrature, the uncorrelated, total uncertainty on $E(\eta)$ is found
to be $6.1\%$.  Systematic uncertainties can in general \emph{not} be
cancelled between the numerator and denominator of \eqref{eq:empcor},
since the same $\eta$ regions are probed by different detector
elements in each.

Note, that the previously published result~\cite{Abbas:2013bpa} used
in \eqref{eq:empcor} already carries a $2\%$ systematic uncertainty
from the particle composition and $p_{\text{T}}$ distribution
\cite{Abbas:2013bpa}. This contribution is contained in the $4.7\%$
quoted above, and is propagated to the final $6.1\%$ systematic
uncertainty on $E(\eta)$.

Finally, it was found through simulations that the acceptance region
of FMD1 is particularly affected by the variations in the number of
secondary particles stemming from variations in the
\systUncName{particle composition and $p_{\text{T}}$} distribution,
and gives rise to an additional $2\%$ systematic uncertainty, which is
added in quadrature to the rest of the systematic uncertainties, but
only for $\eta>3.7$.

\begin{table}[htbp]
  \centering
  \begingroup
  \small 
  \begin{tabular}[t]{llc}
    \hline 
    \textbf{Detector} 
    & \multicolumn{1}{c}{\textbf{Source}}
    & \textbf{Uncertainty} 
    \\
    & 
    & \textbf{(\%)}\\ 
    \hline 
    Common 
    & Centrality 
    & $0.4 - 6.2$ 
    \\ 
    \hline 
    SPD 
    & Background subtraction 
    & $0.1$
    \\
    & Particle composition 
    & $1$ 
    \\ 
    & Weak decays 
    & $1$ 
    \\ 
    & Extrapolation to $p_{\text{T}}=0$ 
    & $2$
    \\ 
    & Event generator 
    & $2$ 
    \\
    & Acceptance 
    & $0-2^{\dag}$
    \\
    \hline 
    FMD 
    & Recombination 
    & $1$
    \\ 
    & Threshold 
    & ${}^{+1}_{-2}$ 
    \\
    & Secondary particles 
    & $6.1$
    \\ 
    & Particle composition \& $p_{\text{T}}$ 
    & $2^{\ddag}$
    \\ 
    \hline 
  \end{tabular}
  \endgroup
  \caption{Summary of systematic uncertainties: the common systematic
    uncertainties shared by both the SPD and the FMD, and the uncertainties
    particular to the detectors.  ${}^{\dag}$Pseudorapidity{}
    dependent uncertainty, largest at $|\eta|=2$. ${}^{\ddag}$Additional
    contribution in $3.7<\eta<5$. See also text.} 
  \label{tab:sysuncer}
\end{table}


\section{Results} 
\label{sec:results}

\begin{figure}[htbp]
  \centering
  \figinput[\linewidth]{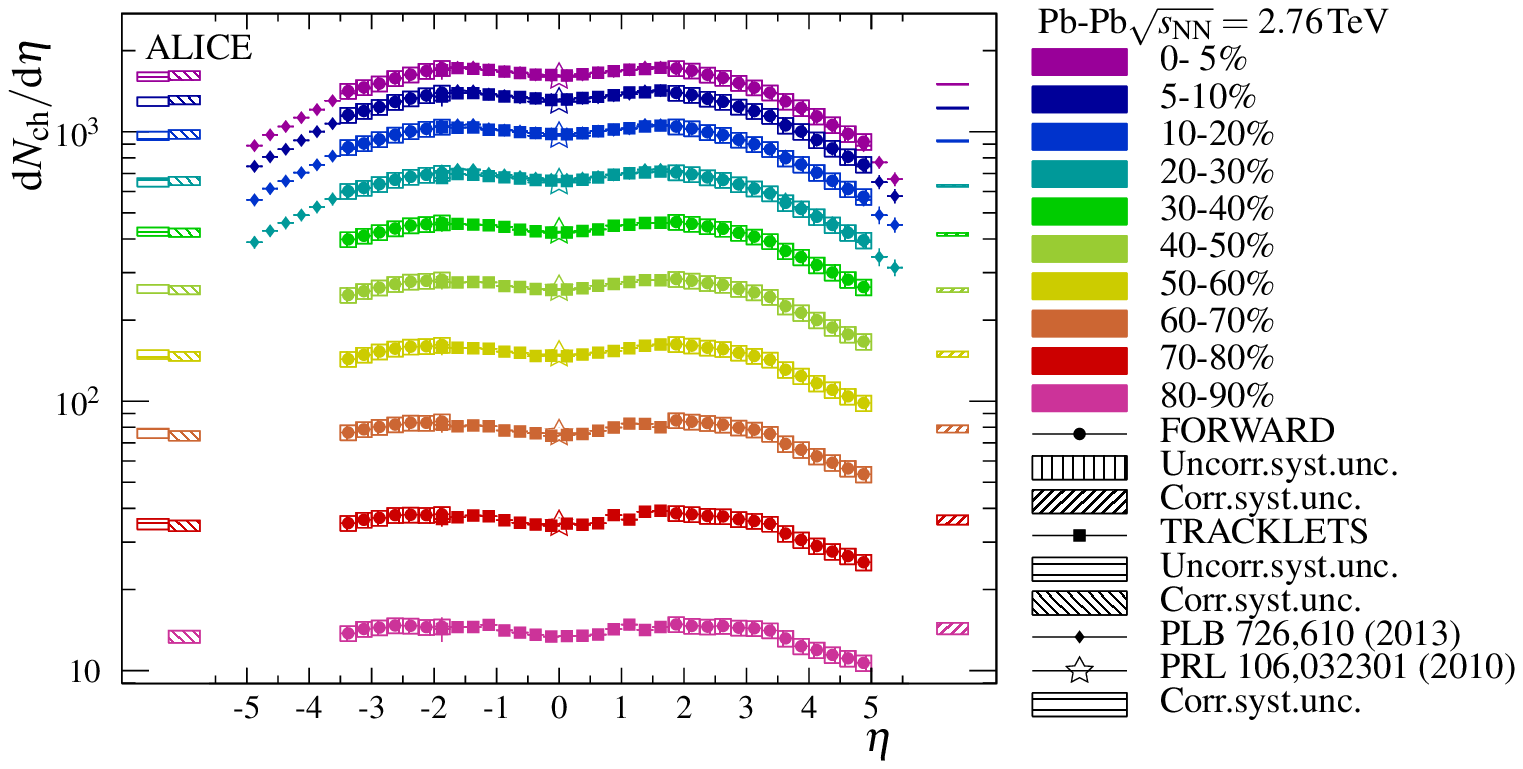}
  \input{figs/caps/PerformanceLogy.cap}
  \label{fig:performancelogy}
\end{figure}

\figref*{fig:performancelogy} shows the charged--particle pseudorapidity
density for different centralities from each detector separately.  

\begin{figure*}[htbp]
  \centering
  \figinput[\linewidth]{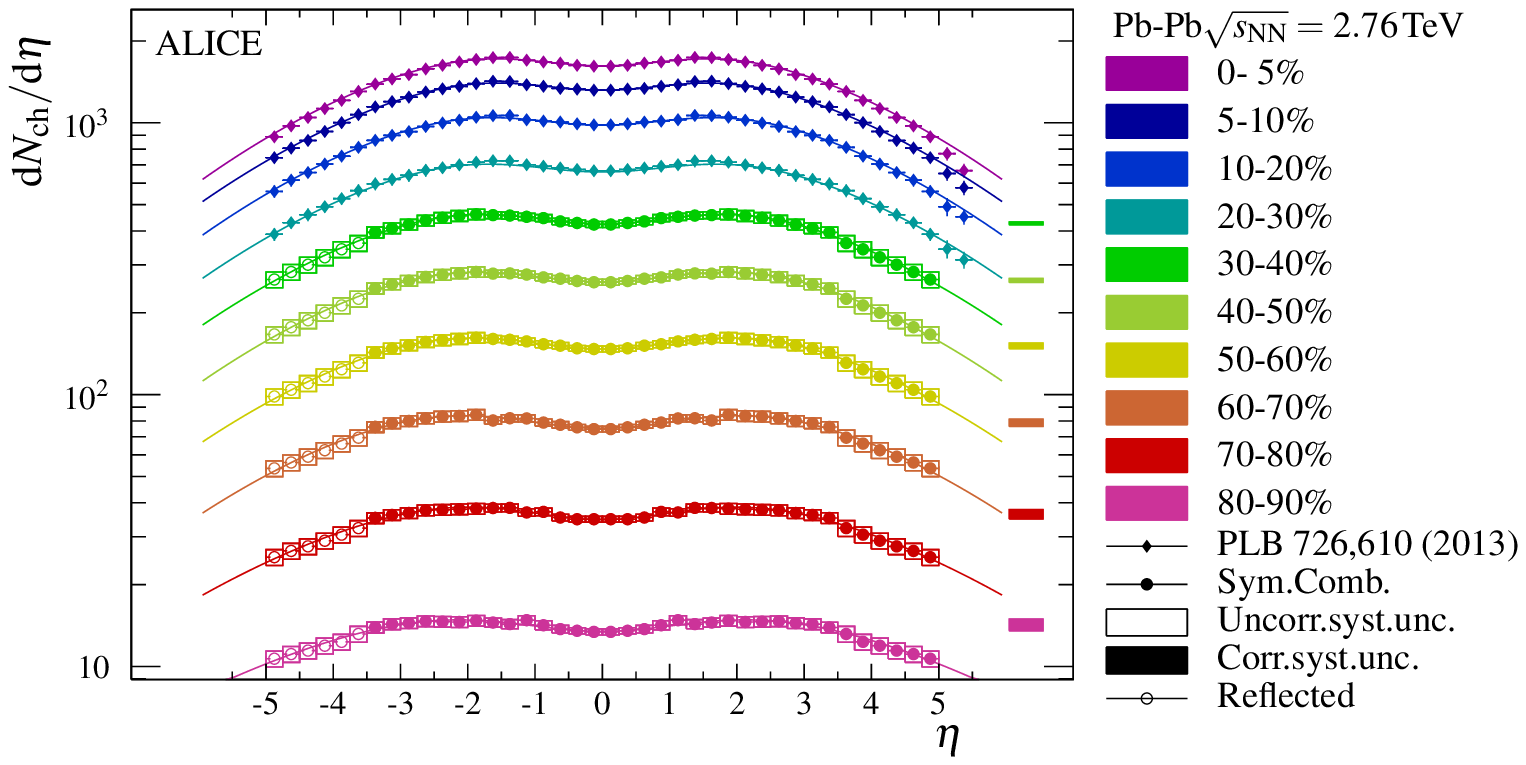}
  \input{figs/caps/ResultLogy.cap}
  \label{fig:resultlogy}
\end{figure*}

The combined distributions in \figref{fig:resultlogy} are calculated
as the average of the individual measurements from the FMD and SPD,
weighted by statistical errors and systematic uncertainties, omitting
those which are common such as that from the centrality determination.
The distributions are then symmetrised around $\eta=0$ by taking the
weighted average of $\pm\eta$ points.  Points at $3.5<\eta<5$ are
reflected on to $-5<\eta<-3.5$ to provide the $\ndndeta$ distributions
in a range comparable to the previously published
results~\cite{Abbas:2013bpa}.

The lines in \figref{fig:resultlogy} are fits of 
\begin{equation}
  \label{eq:gminusg}
  f_{\text{GG}}(\eta;A_1,\sigma_1,A_2,\sigma_2) =
  A_1e^{-\frac12\frac{\eta^2}{\sigma_1^2}} -
  A_2e^{-\frac12\frac{\eta^2}{\sigma_2^2}}\quad,
\end{equation}
to the measured distributions.  The function $f_{\text{GG}}$ is the
difference of two Gaussian distributions centred at $\eta=0$ with
amplitudes $A_1$, $A_2$, and widths $\sigma_1$, $\sigma_2$.  The
function describes the data well within the measured region with a
reduced $\chi^2$ smaller than 1. We find values of $A_2/A_1$ for all
centralities, from $0.20$ to $0.31$ but are consistent within fit
uncertainties, with a constant value of $0.23\pm0.02$.  Likewise
values of $\sigma_2/\sigma_1$ for all centralities, ranges from $0.28$
to $0.36$ and are consistent with a constant value of $0.31\pm0.02$.

\begin{figure}[htbp]
  \centering
  \figinput[\linewidth]{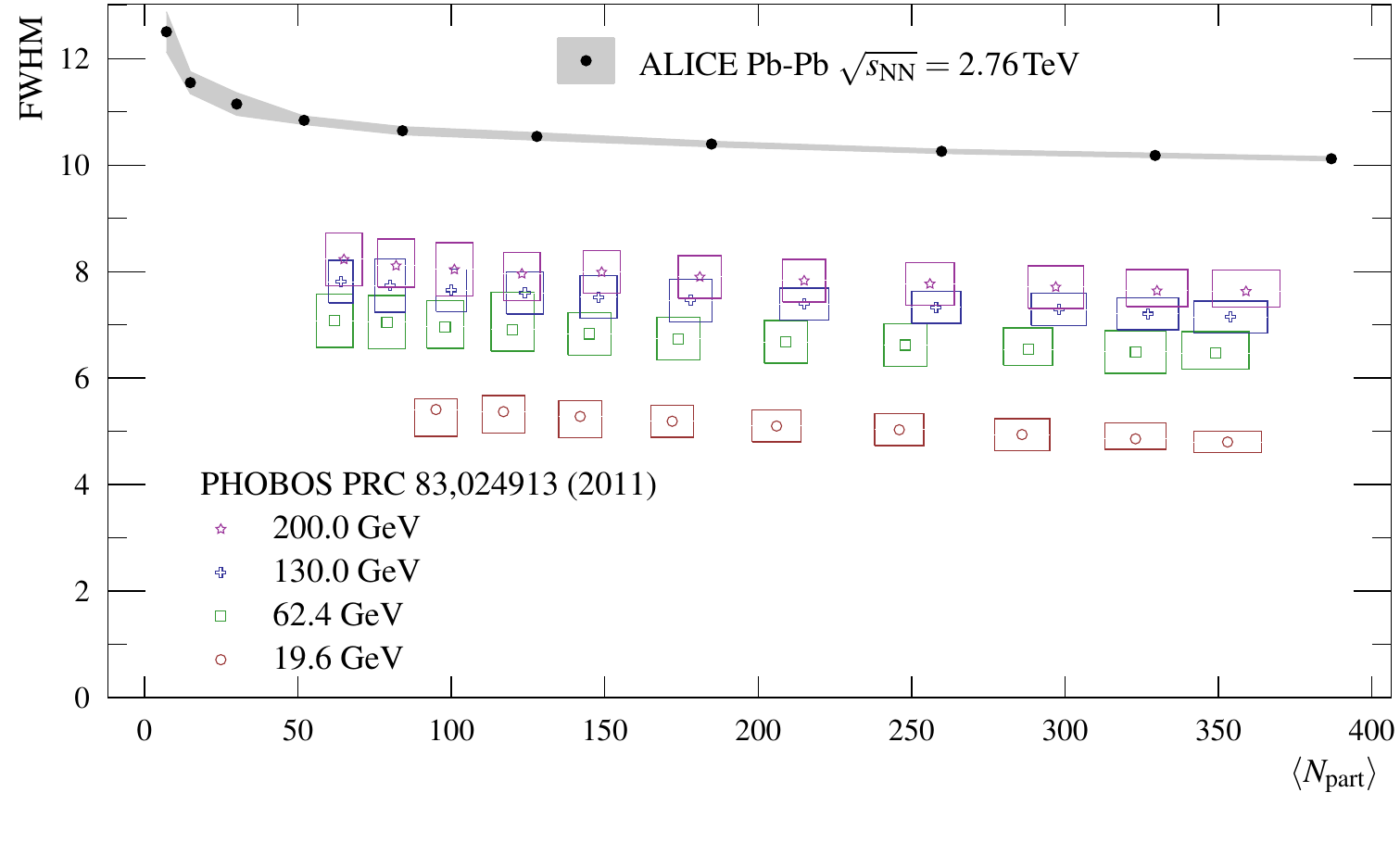}
  \input{figs/caps/FWHM.cap}
  \label{fig:fwhm}
\end{figure}

Qualitatively the shape of the charged--particle pseudorapidity density
distributions broadens only slightly toward more peripheral events,
consistent with the above observation.  Indeed, the full--width
half-maximum (FWHM) shown in~\figref{fig:fwhm} versus the number of
participating nucleons $\langle N_{\text{part}}\rangle$ --- calculated
using a Glauber model~\cite{Abelev:2013qoq} --- increase sharply only
in the very most peripheral collisions.  The $\ndndeta$ distributions
does not extend far enough to calculate reliable values for FWHM
directly from the data.  Instead
$f_{\text{GG}}(\eta)-\max(f_{\text{GG}})/2=0$ was numerically solved,
and the uncertainties evaluated as the error of $f_{\text{GG}}$ at the
roots, divided by the slope at those roots.  The width of the
$\ndndeta$ distributions follows the same trend, in the region of
\centRange{0}{50}, as was seen in lower energy results from PHOBOS
reproduced in \figref*{fig:fwhm} for comparison
\cite{PhysRevC.83.024913}.

\begin{figure}[htbp]
  \centering
  \figinput[\linewidth]{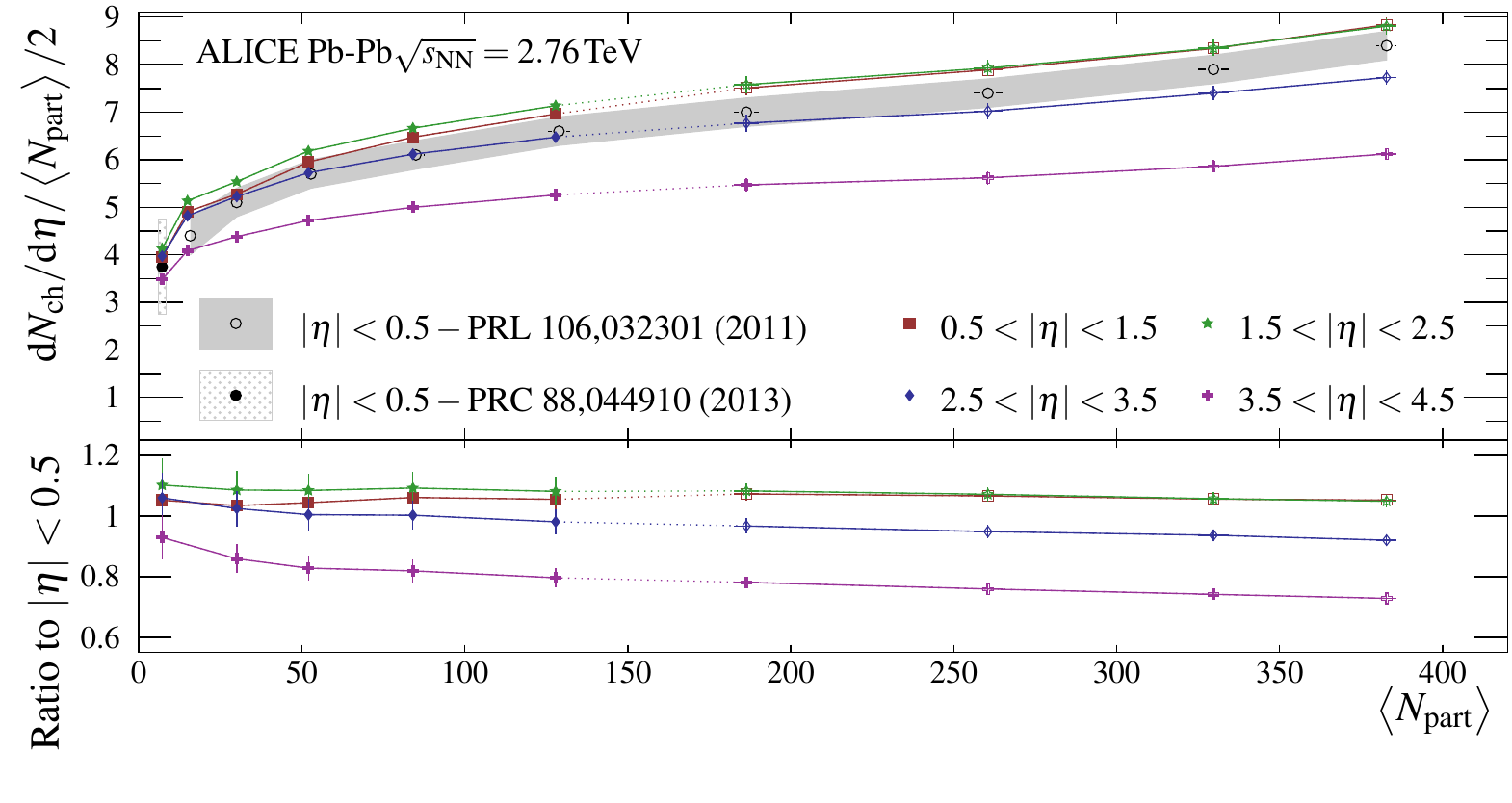}  
  \input{figs/caps/NpartEta.cap}
  \label{fig:nparteta}
\end{figure}

\figref*{fig:nparteta} presents the charged--particle pseudorapidity density
per average number of participating nucleon pairs
$\langle N_{\text{part}}\rangle/2$ as a function of the average number
of participants $\langle N_{\text{part}}\rangle$.  Although there is a
slight increase in the ratio to the central pseudorapidity density
distribution at low $\langle N_{\text{part}}\rangle$ (see lower part
of \figref{fig:nparteta}), the uncertainties are large and no strong
evolution of the shape of the pseudorapidity density distribution over pseudorapidity
with respect to centrality is observed.  The ratio at $3.5<|\eta|<4.5$
does deviate somewhat in peripheral collisions, which is attributed to the
general broadening of the pseudorapidity density distributions in those
collisions. 

To extract the total number of charged particles produced in Pb--Pb
collisions at various centralities, a number of functions, including
\eqref{eq:gminusg}, is fitted to the $\ndndeta$ distributions. A
trapezoid
\begin{align}
  \label{eq:trap}
  f_{\text{T}}(\eta;y_{\text{beam}},M,A) 
  &= A\times\begin{cases}
    0 & |\eta| > y_{\text{beam}}\\
    (y_{\text{beam}}+\eta) & \eta < -M\\
    (y_{\text{beam}}-M) & |\eta|<M\\
    (y_{\text{beam}}-\eta) & \eta > +M
  \end{cases}\quad,
\end{align}
was successfully used by PHOBOS to describe limiting fragmentation
\cite{PhysRevC.83.024913}. Here, $[-M,M]$ is the range in which the
function is constant, and $A$ is the amplitude.  The
parameterisation
\begin{align}
  \label{eq:phobos}
  f_{\text{P}}(\eta;A,\alpha,\beta,a) 
  &= A\frac{\sqrt{1-1/\left[\alpha\cosh(\eta)\right]^2}}{
    1+e^{(|\eta|-\beta)/a}}\quad, 
\end{align}
as suggested by PHOBOS, is likewise fitted to the $\ndndeta$
distributions.  The parameter $a$ expresses the width of the
distribution, and $\alpha$ and $\beta$, and expresses the width and
depth of the dip at $\eta\approx0$, respectively. $A$ is an overall
scale parameter.  Finally, to remedy some of the obvious defects of
the trapezoid i.e., a non--continuous first derivative at $\eta=M$, we
use a Bjorken--inspired function~\cite{Abbas:2013bpa}
\begin{align}
  \label{eq:bjorken}
  f_{\text{B}}(\eta;A,\mu,\sigma) 
  &= A\times\begin{cases} 
    e^{-\frac{(\eta+\mu)^2}{2\sigma^2}} & \eta < -\mu\\
    e^{-\frac{(\eta-\mu)^2}{2\sigma^2}} & \eta > +\mu\\
    1 & |\eta|<\mu
  \end{cases}\quad,        
\end{align}
which has plateau at $A$ for $|\eta|<\mu$ connected to Gaussian
fall--off beyond $\pm\mu$.  The fitted functions are integrated over
$\eta$ up to the beam rapidity $\pm y_{\text{beam}}=\pm7.99$.
Although the $\ndndeta$ distributions in principle continue to
infinity, there is no significant loss in generality or precision by
cutting the integral at $\eta=\pm y_{\text{beam}}$ since the
distributions rapidly approach zero.  Notice that all parameters of
the functions are left free in the fitting procedure.  All functions
give reasonable fits (with a reduced $\chi^2$ smaller than 1), though
the trapezoid and Bjorken--inspired ansatz are too flat at the
mid--rapidity.  The calculation of the central values and uncertainties
are done as for previous results \cite{Abbas:2013bpa}: The central
value is calculated from the integral of the trapezoid fit to compare
directly to previous results; the spread between the integrals and the
central value is evaluated to obtain the uncertainty on the total
$N_{\text{ch}}$.

\begin{figure}[htbp]
  \centering
  {\bgroup\def\myMarkSize{4pt}
    \figinput[\linewidth]{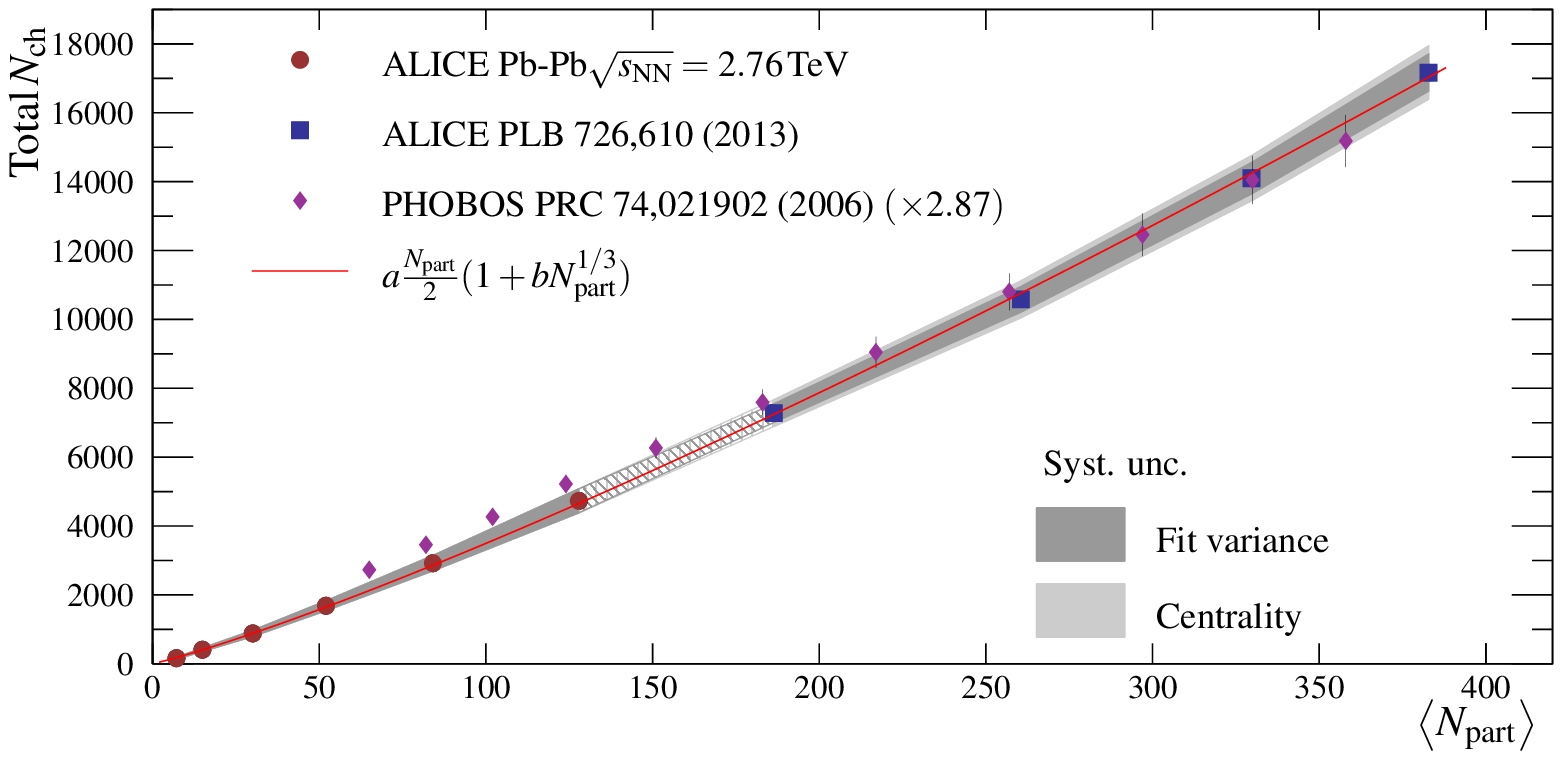}
    \egroup}
  \input{figs/caps/FitsNpart.cap}
  \label{fig:fitsnpart}
\end{figure}

The extrapolated total $N_{\text{ch}}$ versus
$\langle N_{\text{part}}\rangle$ is shown in \figref{fig:fitsnpart},
and compared to lower energy results from PHOBOS~\cite{Alver:2008ck}.
At LHC energies the particle production as a function of
$\langle N_{\text{part}}\rangle$ shows a similar behaviour to the
lower energy results, and the factorisation~\cite{PhysRevC.83.024913}
in centrality and energy seems to hold (see fit in
\figref{fig:fitsnpart}).

\begin{figure}[htbp]
  \centering
  {\bgroup\def\myThickLine{1.8}
    \figinput[\linewidth]{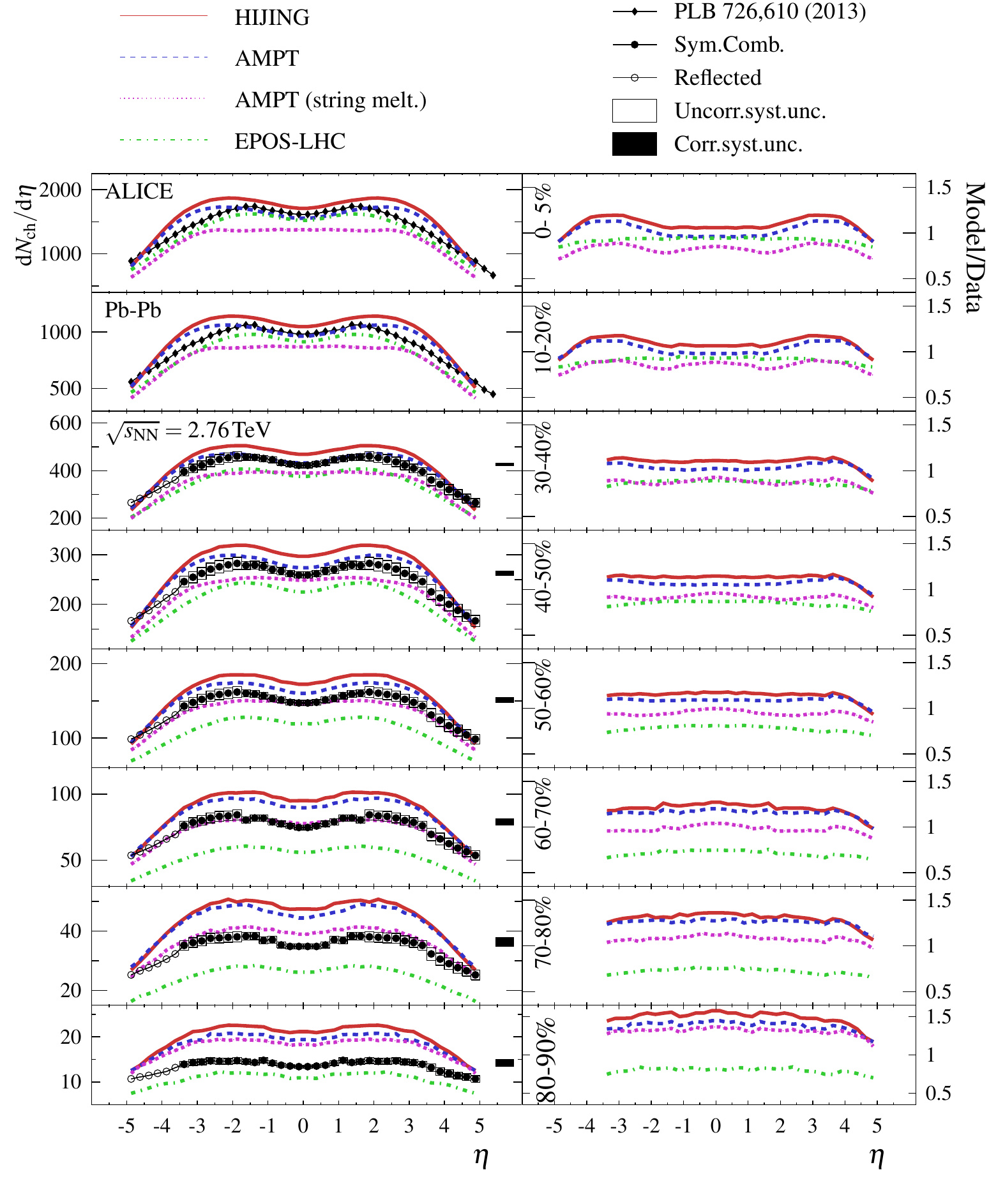}
    \egroup}
  \input{figs/caps/Models8.cap}
  \label{fig:Models}
\end{figure}

In \figref{fig:Models} we show comparisons of various model
calculations to the measured charged--particle pseudorapidity density as a
function of centrality.  The centrality class for a given
model--generated event was determined by sharp cuts in the impact
parameter $b$ and a Glauber calculation \cite{Abelev:2013qoq}.

The HIJING model \cite{Wang:1991hta} (version 1.383, with
jet--quenching disabled, shadowing enabled, and a hard $p_{\text{T}}$
cut--off of \unit[2.3]{Ge\kern-.2exV}) is seen to overshoot the data
for all centralities. In addition, the distributions at all
centralities decrease with increasing $|\eta|$ faster than the data
would suggest.

AMPT~\cite{Lin:2004en} without string melting reproduces the data
fairly well at central pseudorapidity for the most central events --- exactly
in the region it was tuned to, but it fails to describe the
charged--particle pseudorapidity density for more peripheral events.  Also,
AMPT without string melting would suggest a wider central region than
supported by data, and similarly to HIJING decreases faster than the
data.  AMPT with string melting --- which essentially implements quark
coalescence, and therefore a more predominant parton phase --- is seen
to be very flat at mid--rapidity and underestimates the yield, except for
peripheral collisions.

Finally, EPOS--LHC~\cite{Pierog:2013ria} reproduces the shape fairly
well, but underestimates the data by 10 to 30\%.



\section{Conclusions}
\label{sec:concl}

The charged--particle pseudorapidity density has been measured in Pb--Pb
collisions at $\sNN$ over a broad pseudorapidity range, extending previous
published results by ALICE to more peripheral collisions.  
In the mid--rapidity region the well--established tracklet procedure was
used.  In the forward regions, a new data--driven procedure to correct
for the large background due to secondary particles was used.  
The results presented here are consistent with the behaviour
previously seen in more central collisions and in a limited pseudorapidity
range.  
No strong evolution of the overall shape of the charged--particle
pseudorapidity density distributions as a function of collision centrality is
observed.  When normalised to the number of participating nucleons in
the collision, the centrality evolution is small over the pseudorapidity
range.  
Since the measurement was performed over a large pseudorapidity range
($-3.5<\eta<5$), it allows for an estimate of the total number of
charged particles produced in Pb--Pb collisions at $\sNN$.  The total
charged--particle multiplicity is found to scale approximately with
the number of participating nucleons.  This would suggest that hard
contributions to the total charged--particle multiplicity are small.
From peripheral to central collisions we observe an increase of two
orders of magnitude in the number of produced charge particles. 
A comparison of the data to the different available predictions from
HIJING, AMPT, and EPOS-LHC show that none of these models captures
both the shape and level of the measured distributions.  
AMPT however comes close in limited ranges of centrality.  The exact
centrality ranges that AMPT describes depend strongly on whether
string melting is used in the model or not.  EPOS-LHC --- although
systematically low --- shows a reasonable agreement with the shape of
the measured charged--particle pseudorapidity density distribution over a
wider pseudorapidity range.



\section*{Acknowledgements}
\label{app:acknowledgements}

The ALICE Collaboration would like to thank all its engineers and technicians for their invaluable contributions to the construction of the experiment and the CERN accelerator teams for the outstanding performance of the LHC complex.
The ALICE Collaboration gratefully acknowledges the resources and support provided by all Grid centres and the Worldwide LHC Computing Grid (WLCG) collaboration.
The ALICE Collaboration acknowledges the following funding agencies for their support in building and
running the ALICE detector:
State Committee of Science,  World Federation of Scientists (WFS)
and Swiss Fonds Kidagan, Armenia;
Conselho Nacional de Desenvolvimento Cient\'{\i}fico e Tecnol\'{o}gico (CNPq), Financiadora de Estudos e Projetos (FINEP),
Funda\c{c}\~{a}o de Amparo \`{a} Pesquisa do Estado de S\~{a}o Paulo (FAPESP);
National Natural Science Foundation of China (NSFC), the Chinese Ministry of Education (CMOE)
and the Ministry of Science and Technology of China (MSTC);
Ministry of Education and Youth of the Czech Republic;
Danish Natural Science Research Council, the Carlsberg Foundation and the Danish National Research Foundation;
The European Research Council under the European Community's Seventh Framework Programme;
Helsinki Institute of Physics and the Academy of Finland;
French CNRS-IN2P3, the `Region Pays de Loire', `Region Alsace', `Region Auvergne' and CEA, France;
German Bundesministerium fur Bildung, Wissenschaft, Forschung und Technologie (BMBF) and the Helmholtz Association;
General Secretariat for Research and Technology, Ministry of Development, Greece;
Hungarian Orszagos Tudomanyos Kutatasi Alappgrammok (OTKA) and National Office for Research and Technology (NKTH);
Department of Atomic Energy and Department of Science and Technology of the Government of India;
Istituto Nazionale di Fisica Nucleare (INFN) and Centro Fermi -
Museo Storico della Fisica e Centro Studi e Ricerche ``Enrico Fermi'', Italy;
MEXT Grant-in-Aid for Specially Promoted Research, Ja\-pan;
Joint Institute for Nuclear Research, Dubna;
National Research Foundation of Korea (NRF);
Consejo Nacional de Cienca y Tecnologia (CONACYT), Direccion General de Asuntos del Personal Academico(DGAPA), M\'{e}xico, Amerique Latine Formation academique - 
European Commission~(ALFA-EC) and the EPLANET Program~(European Particle Physics Latin American Network);
Stichting voor Fundamenteel Onderzoek der Materie (FOM) and the Nederlandse Organisatie voor Wetenschappelijk Onderzoek (NWO), Netherlands;
Research Council of Norway (NFR);
National Science Centre, Poland;
Ministry of National Education/Institute for Atomic Physics and National Council of Scientific Research in Higher Education~(CNCSI-UEFISCDI), Romania;
Ministry of Education and Science of Russian Federation, Russian
Academy of Sciences, Russian Federal Agency of Atomic Energy,
Russian Federal Agency for Science and Innovations and The Russian
Foundation for Basic Research;
Ministry of Education of Slovakia;
Department of Science and Technology, South Africa;
Centro de Investigaciones Energeticas, Medioambientales y Tecnologicas (CIEMAT), E-Infrastructure shared between Europe and Latin America (EELA), 
Ministerio de Econom\'{i}a y Competitividad (MINECO) of Spain, Xunta de Galicia (Conseller\'{\i}a de Educaci\'{o}n),
Centro de Aplicaciones Tecnológicas y Desarrollo Nuclear (CEA\-DEN), Cubaenerg\'{\i}a, Cuba, and IAEA (International Atomic Energy Agency);
Swedish Research Council (VR) and Knut $\&$ Alice Wallenberg
Foundation (KAW);
Ukraine Ministry of Education and Science;
United Kingdom Science and Technology Facilities Council (STFC);
The United States Department of Energy, the United States National
Science Foundation, the State of Texas, and the State of Ohio;
Ministry of Science, Education and Sports of Croatia and  Unity through Knowledge Fund, Croatia;
Council of Scientific and Industrial Research (CSIR), New Delhi, India;
Pontificia Universidad Cat\'{o}lica del Per\'{u}.
 
\mbox{\hfill}

\bibliography{bib}

\clearpage
\appendix 
\section{The ALICE Collaboration}
\label{app:collab}



\begingroup
\small
\begin{flushleft}
J.~Adam\Irefn{org40}\And
D.~Adamov\'{a}\Irefn{org83}\And
M.M.~Aggarwal\Irefn{org87}\And
G.~Aglieri Rinella\Irefn{org36}\And
M.~Agnello\Irefn{org110}\And
N.~Agrawal\Irefn{org48}\And
Z.~Ahammed\Irefn{org132}\And
S.U.~Ahn\Irefn{org68}\And
S.~Aiola\Irefn{org136}\And
A.~Akindinov\Irefn{org58}
% ...
\renewcommand\labelenumi{\textsuperscript{\theenumi}~}

\section*{Affiliation notes}
\renewcommand\theenumi{\roman{enumi}}
\begin{Authlist}
\item \Adef{0}Deceased
\item \Adef{idp1749136}{Also at: Georgia State University, Atlanta, Georgia, United States}
\item \Adef{idp3785536}{Also at: M.V. Lomonosov Moscow State University, D.V. Skobeltsyn Institute of Nuclear, Physics, Moscow, Russia}
\end{Authlist}

\section*{Collaboration Institutes}
\renewcommand\theenumi{\arabic{enumi}~}
\begin{Authlist}

\item \Idef{org1}A.I. Alikhanyan National Science Laboratory (Yerevan Physics Institute) Foundation, Yerevan, Armenia
\item \Idef{org2}Benem\'{e}rita Universidad Aut\'{o}noma de Puebla, Puebla, Mexico
\item \Idef{org3}Bogolyubov Institute for Theoretical Physics, Kiev, Ukraine
\item \Idef{org4}Bose Institute, Department of Physics and Centre for Astroparticle Physics and Space Science (CAPSS), Kolkata, India
\item \Idef{org5}Budker Institute for Nuclear Physics, Novosibirsk, Russia
\item \Idef{org6}California Polytechnic State University, San Luis Obispo, California, United States
\item \Idef{org7}Central China Normal University, Wuhan, China
\item \Idef{org8}Centre de Calcul de l'IN2P3, Villeurbanne, France
\item \Idef{org9}Centro de Aplicaciones Tecnol\'{o}gicas y Desarrollo Nuclear (CEADEN), Havana, Cuba
\item \Idef{org10}Centro de Investigaciones Energ\'{e}ticas Medioambientales y Tecnol\'{o}gicas (CIEMAT), Madrid, Spain
\item \Idef{org11}Centro de Investigaci\'{o}n y de Estudios Avanzados (CINVESTAV), Mexico City and M\'{e}rida, Mexico
\item \Idef{org12}Centro Fermi - Museo Storico della Fisica e Centro Studi e Ricerche ``Enrico Fermi'', Rome, Italy
\item \Idef{org13}Chicago State University, Chicago, Illinois, USA
\item \Idef{org14}China Institute of Atomic Energy, Beijing, China
\item \Idef{org15}Commissariat \`{a} l'Energie Atomique, IRFU, Saclay, France
\item \Idef{org16}COMSATS Institute of Information Technology (CIIT), Islamabad, Pakistan
\item \Idef{org17}Departamento de F\'{\i}sica de Part\'{\i}culas and IGFAE, Universidad de Santiago de Compostela, Santiago de Compostela, Spain
\item \Idef{org18}Department of Physics and Technology, University of Bergen, Bergen, Norway
\item \Idef{org19}Department of Physics, Aligarh Muslim University, Aligarh, India
\item \Idef{org20}Department of Physics, Ohio State University, Columbus, Ohio, United States
\item \Idef{org21}Department of Physics, Sejong University, Seoul, South Korea
\item \Idef{org22}Department of Physics, University of Oslo, Oslo, Norway
\item \Idef{org23}Dipartimento di Elettrotecnica ed Elettronica del Politecnico, Bari, Italy
\item \Idef{org24}Dipartimento di Fisica dell'Universit\`{a} 'La Sapienza' and Sezione INFN Rome, Italy
\item \Idef{org25}Dipartimento di Fisica dell'Universit\`{a} and Sezione INFN, Cagliari, Italy
\item \Idef{org26}Dipartimento di Fisica dell'Universit\`{a} and Sezione INFN, Trieste, Italy
\item \Idef{org27}Dipartimento di Fisica dell'Universit\`{a} and Sezione INFN, Turin, Italy
\item \Idef{org28}Dipartimento di Fisica e Astronomia dell'Universit\`{a} and Sezione INFN, Bologna, Italy
\item \Idef{org29}Dipartimento di Fisica e Astronomia dell'Universit\`{a} and Sezione INFN, Catania, Italy
\item \Idef{org30}Dipartimento di Fisica e Astronomia dell'Universit\`{a} and Sezione INFN, Padova, Italy
\item \Idef{org31}Dipartimento di Fisica `E.R.~Caianiello' dell'Universit\`{a} and Gruppo Collegato INFN, Salerno, Italy
\item \Idef{org32}Dipartimento di Scienze e Innovazione Tecnologica dell'Universit\`{a} del  Piemonte Orientale and Gruppo Collegato INFN, Alessandria, Italy
\item \Idef{org33}Dipartimento Interateneo di Fisica `M.~Merlin' and Sezione INFN, Bari, Italy
\item \Idef{org34}Division of Experimental High Energy Physics, University of Lund, Lund, Sweden
\item \Idef{org35}Eberhard Karls Universit\"{a}t T\"{u}bingen, T\"{u}bingen, Germany
\item \Idef{org36}European Organization for Nuclear Research (CERN), Geneva, Switzerland
\item \Idef{org37}Excellence Cluster Universe, Technische Universit\"{a}t M\"{u}nchen, Munich, Germany
\item \Idef{org38}Faculty of Engineering, Bergen University College, Bergen, Norway
\item \Idef{org39}Faculty of Mathematics, Physics and Informatics, Comenius University, Bratislava, Slovakia
\item \Idef{org40}Faculty of Nuclear Sciences and Physical Engineering, Czech Technical University in Prague, Prague, Czech Republic
\item \Idef{org41}Faculty of Science, P.J.~\v{S}af\'{a}rik University, Ko\v{s}ice, Slovakia
\item \Idef{org42}Faculty of Technology, Buskerud and Vestfold University College, Vestfold, Norway
\item \Idef{org43}Frankfurt Institute for Advanced Studies, Johann Wolfgang Goethe-Universit\"{a}t Frankfurt, Frankfurt, Germany
\item \Idef{org44}Gangneung-Wonju National University, Gangneung, South Korea
\item \Idef{org45}Gauhati University, Department of Physics, Guwahati, India
\item \Idef{org46}Helsinki Institute of Physics (HIP), Helsinki, Finland
\item \Idef{org47}Hiroshima University, Hiroshima, Japan
\item \Idef{org48}Indian Institute of Technology Bombay (IIT), Mumbai, India
\item \Idef{org49}Indian Institute of Technology Indore, Indore (IITI), India
\item \Idef{org50}Inha University, Incheon, South Korea
\item \Idef{org51}Institut de Physique Nucl\'eaire d'Orsay (IPNO), Universit\'e Paris-Sud, CNRS-IN2P3, Orsay, France
\item \Idef{org52}Institut f\"{u}r Informatik, Johann Wolfgang Goethe-Universit\"{a}t Frankfurt, Frankfurt, Germany
\item \Idef{org53}Institut f\"{u}r Kernphysik, Johann Wolfgang Goethe-Universit\"{a}t Frankfurt, Frankfurt, Germany
\item \Idef{org54}Institut f\"{u}r Kernphysik, Westf\"{a}lische Wilhelms-Universit\"{a}t M\"{u}nster, M\"{u}nster, Germany
\item \Idef{org55}Institut Pluridisciplinaire Hubert Curien (IPHC), Universit\'{e} de Strasbourg, CNRS-IN2P3, Strasbourg, France
\item \Idef{org56}Institute for Nuclear Research, Academy of Sciences, Moscow, Russia
\item \Idef{org57}Institute for Subatomic Physics of Utrecht University, Utrecht, Netherlands
\item \Idef{org58}Institute for Theoretical and Experimental Physics, Moscow, Russia
\item \Idef{org59}Institute of Experimental Physics, Slovak Academy of Sciences, Ko\v{s}ice, Slovakia
\item \Idef{org60}Institute of Physics, Academy of Sciences of the Czech Republic, Prague, Czech Republic
\item \Idef{org61}Institute of Physics, Bhubaneswar, India
\item \Idef{org62}Institute of Space Science (ISS), Bucharest, Romania
\item \Idef{org63}Instituto de Ciencias Nucleares, Universidad Nacional Aut\'{o}noma de M\'{e}xico, Mexico City, Mexico
\item \Idef{org64}Instituto de F\'{\i}sica, Universidad Nacional Aut\'{o}noma de M\'{e}xico, Mexico City, Mexico
\item \Idef{org65}iThemba LABS, National Research Foundation, Somerset West, South Africa
\item \Idef{org66}Joint Institute for Nuclear Research (JINR), Dubna, Russia
\item \Idef{org67}Konkuk University, Seoul, South Korea
\item \Idef{org68}Korea Institute of Science and Technology Information, Daejeon, South Korea
\item \Idef{org69}KTO Karatay University, Konya, Turkey
\item \Idef{org70}Laboratoire de Physique Corpusculaire (LPC), Clermont Universit\'{e}, Universit\'{e} Blaise Pascal, CNRS--IN2P3, Clermont-Ferrand, France
\item \Idef{org71}Laboratoire de Physique Subatomique et de Cosmologie, Universit\'{e} Grenoble-Alpes, CNRS-IN2P3, Grenoble, France
\item \Idef{org72}Laboratori Nazionali di Frascati, INFN, Frascati, Italy
\item \Idef{org73}Laboratori Nazionali di Legnaro, INFN, Legnaro, Italy
\item \Idef{org74}Lawrence Berkeley National Laboratory, Berkeley, California, United States
\item \Idef{org75}Moscow Engineering Physics Institute, Moscow, Russia
\item \Idef{org76}Nagasaki Institute of Applied Science, Nagasaki, Japan
\item \Idef{org77}National Centre for Nuclear Studies, Warsaw, Poland
\item \Idef{org78}National Institute for Physics and Nuclear Engineering, Bucharest, Romania
\item \Idef{org79}National Institute of Science Education and Research, Bhubaneswar, India
\item \Idef{org80}Niels Bohr Institute, University of Copenhagen, Copenhagen, Denmark
\item \Idef{org81}Nikhef, Nationaal instituut voor subatomaire fysica, Amsterdam, Netherlands
\item \Idef{org82}Nuclear Physics Group, STFC Daresbury Laboratory, Daresbury, United Kingdom
\item \Idef{org83}Nuclear Physics Institute, Academy of Sciences of the Czech Republic, \v{R}e\v{z} u Prahy, Czech Republic
\item \Idef{org84}Oak Ridge National Laboratory, Oak Ridge, Tennessee, United States
\item \Idef{org85}Petersburg Nuclear Physics Institute, Gatchina, Russia
\item \Idef{org86}Physics Department, Creighton University, Omaha, Nebraska, United States
\item \Idef{org87}Physics Department, Panjab University, Chandigarh, India
\item \Idef{org88}Physics Department, University of Athens, Athens, Greece
\item \Idef{org89}Physics Department, University of Cape Town, Cape Town, South Africa
\item \Idef{org90}Physics Department, University of Jammu, Jammu, India
\item \Idef{org91}Physics Department, University of Rajasthan, Jaipur, India
\item \Idef{org92}Physik Department, Technische Universit\"{a}t M\"{u}nchen, Munich, Germany
\item \Idef{org93}Physikalisches Institut, Ruprecht-Karls-Universit\"{a}t Heidelberg, Heidelberg, Germany
\item \Idef{org94}Purdue University, West Lafayette, Indiana, United States
\item \Idef{org95}Pusan National University, Pusan, South Korea
\item \Idef{org96}Research Division and ExtreMe Matter Institute EMMI, GSI Helmholtzzentrum f\"ur Schwerionenforschung, Darmstadt, Germany
\item \Idef{org97}Rudjer Bo\v{s}kovi\'{c} Institute, Zagreb, Croatia
\item \Idef{org98}Russian Federal Nuclear Center (VNIIEF), Sarov, Russia
\item \Idef{org99}Russian Research Centre Kurchatov Institute, Moscow, Russia
\item \Idef{org100}Saha Institute of Nuclear Physics, Kolkata, India
\item \Idef{org101}School of Physics and Astronomy, University of Birmingham, Birmingham, United Kingdom
\item \Idef{org102}Secci\'{o}n F\'{\i}sica, Departamento de Ciencias, Pontificia Universidad Cat\'{o}lica del Per\'{u}, Lima, Peru
\item \Idef{org103}Sezione INFN, Bari, Italy
\item \Idef{org104}Sezione INFN, Bologna, Italy
\item \Idef{org105}Sezione INFN, Cagliari, Italy
\item \Idef{org106}Sezione INFN, Catania, Italy
\item \Idef{org107}Sezione INFN, Padova, Italy
\item \Idef{org108}Sezione INFN, Rome, Italy
\item \Idef{org109}Sezione INFN, Trieste, Italy
\item \Idef{org110}Sezione INFN, Turin, Italy
\item \Idef{org111}SSC IHEP of NRC Kurchatov institute, Protvino, Russia
\item \Idef{org112}Stefan Meyer Institut f\"{u}r Subatomare Physik (SMI), Vienna, Austria
\item \Idef{org113}SUBATECH, Ecole des Mines de Nantes, Universit\'{e} de Nantes, CNRS-IN2P3, Nantes, France
\item \Idef{org114}Suranaree University of Technology, Nakhon Ratchasima, Thailand
\item \Idef{org115}Technical University of Ko\v{s}ice, Ko\v{s}ice, Slovakia
\item \Idef{org116}Technical University of Split FESB, Split, Croatia
\item \Idef{org117}The Henryk Niewodniczanski Institute of Nuclear Physics, Polish Academy of Sciences, Cracow, Poland
\item \Idef{org118}The University of Texas at Austin, Physics Department, Austin, Texas, USA
\item \Idef{org119}Universidad Aut\'{o}noma de Sinaloa, Culiac\'{a}n, Mexico
\item \Idef{org120}Universidade de S\~{a}o Paulo (USP), S\~{a}o Paulo, Brazil
\item \Idef{org121}Universidade Estadual de Campinas (UNICAMP), Campinas, Brazil
\item \Idef{org122}University of Houston, Houston, Texas, United States
\item \Idef{org123}University of Jyv\"{a}skyl\"{a}, Jyv\"{a}skyl\"{a}, Finland
\item \Idef{org124}University of Liverpool, Liverpool, United Kingdom
\item \Idef{org125}University of Tennessee, Knoxville, Tennessee, United States
\item \Idef{org126}University of the Witwatersrand, Johannesburg, South Africa
\item \Idef{org127}University of Tokyo, Tokyo, Japan
\item \Idef{org128}University of Tsukuba, Tsukuba, Japan
\item \Idef{org129}University of Zagreb, Zagreb, Croatia
\item \Idef{org130}Universit\'{e} de Lyon, Universit\'{e} Lyon 1, CNRS/IN2P3, IPN-Lyon, Villeurbanne, France
\item \Idef{org131}V.~Fock Institute for Physics, St. Petersburg State University, St. Petersburg, Russia
\item \Idef{org132}Variable Energy Cyclotron Centre, Kolkata, India
\item \Idef{org133}Warsaw University of Technology, Warsaw, Poland
\item \Idef{org134}Wayne State University, Detroit, Michigan, United States
\item \Idef{org135}Wigner Research Centre for Physics, Hungarian Academy of Sciences, Budapest, Hungary
\item \Idef{org136}Yale University, New Haven, Connecticut, United States
\item \Idef{org137}Yonsei University, Seoul, South Korea
\item \Idef{org138}Zentrum f\"{u}r Technologietransfer und Telekommunikation (ZTT), Fachhochschule Worms, Worms, Germany
\end{Authlist}
\endgroup

\end{document}